\documentstyle[12pt,psfig]{article}
\begin{document}
\newcommand{\beq}{\begin{equation}}
\newcommand{\eeq}{\end{equation}}
\newcommand{\beqa}{\begin{eqnarray}}
\newcommand{\eeqa}{\end{eqnarray}}
\newcommand{\ba}{\begin{array}}
\newcommand{\ea}{\end{array}}

\centerline{\Large \bf Spectral Statistics of the Triaxial Rigid Rotator:}
\centerline{\Large \bf Semiclassical Origin of their Pathological Behavior}
\bigskip
\centerline{
V.R. Manfredi$^{1}$, V. Penna$^{2}$ and L. Salasnich$^{1,3}$}
\bigskip
\begin{center}
$^{1}$Dipartimento di Fisica "G. Galilei", Universit\`a di Padova, \\
Istituto Nazionale di Fisica Nucleare, Sezione di Padova, \\
Via Marzolo 8, 35131 Padova, Italy \\
$^{2}$Dipartimento di Fisica, Politecnico di Torino, \\
Istituto Nazionale per la Fisica della Materia, Unit\`a di Torino, \\
Corso Duca degli Abruzzi 24, 10129 Torino, Italy \\
$^{3}$Dipartimento di Fisica, Universit\`a di Milano, \\ 
Istituto Nazionale per la Fisica della Materia, Unit\`a di Milano, \\
Via Celoria 16, 20133 Milano, Italy
\end{center}
\bigskip

\begin{abstract}
In this paper we investigate the local and global 
spectral properties of the triaxial rigid rotator. 
We demonstrate that, for a fixed value 
of the total angular momentum, 
the energy spectrum can be divided into two sets of energy levels, 
whose classical analog are librational and rotational 
motions. By using diagonalization, semiclassical 
and algebric methods, we show that the energy levels 
follow the anomalous spectral statistics 
of the one-dimensional harmonic oscillator. 
\end{abstract}

\section{Introduction}

Berry and Tabor$^{1,2}$ have shown that for a generic quantum system,  
which is classically integrable, the spectral statistics of energy 
levels follow the Poisson ensemble, 
while for a system whose classical analog is 
chaotic the spectral statistics follow the 
Gaussian Orthogonal Ensemble (GOE) of 
Random Matrix Theory (RMT) (for a review see Ref. 3). 
Nevertheless, there are some exceptions to this classical-quantum 
correspondence: the best known case is the harmonic 
oscillator one.$^{4,5}$ 
Recently we have discussed another pathological case: 
the classically integrable triaxial rigid rotator.$^{6)}$ 
We have numerically computed the energy levels and then
calculated the nearest-neighbor spacing distribution $P(s)$ 
and the spectral rigidity $\Delta_3(L)$. We have found that 
$P(s)$ shows a sharp peak at $s=1$ while $\Delta_3(L)$ for 
small values of $L$ follows the Poissonian predictions 
and asymptotically it shows large fluctuations around 
its mean value.$^{6}$ 
Such behavior of spectral statistics 
is quite similar to that of a one-dimensional harmonic oscillator. 
In this paper we have calculated the exact energy levels, 
the semiclassical ones and those obtained by an algebraic harmonic 
approximation. We show that, 
for a fixed value of the total angular momentum, 
the Hamiltonian of the triaxial top describes a sort of 
nonlinear pendulum and its semiclassical energy spectrum  
can be divided into two sets of energy levels, 
corresponding to librational and rotational motion respectively. 
These energy levels are close to the exact ones 
apart near the separatrix. Around 
the minima and the maxima of the classical energy, 
the exact energy levels are described accurately by 
approximate algebraic formulas.  

\section{Triaxial Rigid Rotator}

The classical Hamiltonian $H$ of the triaxial rigid rotator 
is given by 
\beq 
H = {1 \over 2} \left( a J_1^2 + b J_2^2  + c J_3^2 \right ) \; , 
\label{HAM}
\eeq
where ${\bf J}=(J_1,J_2,J_3)$ is the angular momentum of the 
rotation and $a=1/{\it I_1}$, $b=1/{\it I_2}$, $c=1/{\it I_3}$ 
are three parameters such that ${\it I_1}$, ${\it I_2}$ and ${\it I_3}$ are 
the principal momenta of inertia of the top.$^{7,8}$  
Here we choose $a<b<c$. 
The quantum Hamiltonian ${\hat H}$ is 
obtained by replacing components $J_m$ of the 
angular momentum, in the classical expression (\ref{HAM}), 
by the corresponding quantum operators 
${\hat J}_1$, ${\hat J}_2$ and ${\hat J}_3$.
These obey the commutators 
$[ {\hat J}_{\ell} , {\hat J}_k  ] 
= i \, \varepsilon_{\ell, j , m} {\hat J}_m$ and commute with 
the Casimir operator ${\hat J}^2={\hat J}_1^2+{\hat J}_2^2+{\hat J}_3^2$,
namely $[ {\hat J}^2 , {\hat J}_k ] = 0$. 
In view of the latter equation, a standard choice to deal with 
the angular-momentum dynamics consists in diagonalizing the two 
operators ${\hat J}^2$ and ${\hat J}_3$ within the 
basis $\{ |j,k\rangle: |k| \le j \}$ such that 
${\hat J}_3 |j,k\rangle = k\hbar |j,k\rangle$ and 
${\hat J}^2 |j,k\rangle = j (j +1)\hbar^2 |j,k\rangle$, 
where the index $j$ labels the angular-momentum representation. 
After recalling that 
${\hat J}_{\pm} |j,k\rangle 
= [(j \mp k) (j \pm k+1)]^{1/2} \hbar |j,k \pm 1\rangle$ 
(with ${\hat J}_{\pm}= {\hat J}_{1}\pm i{\hat J}_{2}$), 
the non-zero matrix elements of the quantum Hamiltonian ${\hat H}$ 
in the basis $|j,k\rangle$ are given by 
\beq 
\langle j,k|{\hat H}|j,k \rangle = {\hbar^2 \over 4} (a+b) 
(j(j+1)-k^2) + {\hbar^2 \over 2} c k^2 \; , 
\eeq
$$
\langle j,k|{\hat H}|j,k+2 \rangle = \langle j,k+2|H|j,k \rangle = 
$$ 
\beq
= {\hbar^2 \over 8} (a-b)\sqrt{ (j-k)(j-k-1)(j+k+1)(j+k+2) } \; . 
\eeq 
For a fixed value of $j$, the Hamiltonian matrix can be decomposed 
in four submatrices, corresponding to different classes 
of symmetry.$^{9}$ The numerical diagonalization of the matrix gives 
the "exact" energy levels of triaxial rigid rotator.$^{6}$ 

\section{Semiclassical description}

It is important to observe that, also if the "exact" energy spectrum 
of the triaxial rigid rotator cannot be analytically determined, the 
classical Hamiltonian of the triaxial top is integrable.$^{7,8}$ 
We use the classical integrability of our triaxial top to calculate 
the semiclassical energy spectrum via torus quantization. 
The first step is to find out the action variables of the Hamiltonian. 
By using the Deprit transformation$^{8}$ 
(this is also known, at the quantum level, as Villain's representation 
of angular momentum$^{10}$ 
\beq 
J_1 = \sqrt{J^2 - J_3^2} \; \sin{\theta} \; , 
\;\;\;\;\;\;\; 
J_2 = \sqrt{J^2 - J_3^2} \; \cos{\theta} \; , 
\eeq    
where $J^2=J_1^2+J_2^2+J_3^2$, the classical Hamiltonian can be 
written as  
\beq 
H = {1\over 2} a J^2 + 
{1\over 2} \left[ (c-a) -(b-a)\cos^2{\theta} \right]\, J_3^2 
+ {1\over 2} (b-a) J^2 \cos^2{\theta} \; ,  
\eeq 
where the dynamical variable $J_3$ is the action variable 
of the angle $\theta$.$^{8}$ 
By performing another canonical transformation 
the angular dependence of the Hamiltonian $H=H(J,J_3,\theta)$ 
can be removed. Thus the Hamiltonian 
can be written as a function of two action variables: 
the total angular momentum $J$ and a new action 
variable $I$ given by 
\beq 
I = {J\over \pi} \int_{\theta_1}^{\theta_2}  
\sqrt{ {2 H\over J^2} - a -(b-a)\cos^2{\theta} 
\over (c-a) - (b-a) \cos^2{\theta} } \; d\theta \; . 
\eeq 
The Hamiltonian $H$ is now an implicit function 
of two action variables: $H=H(J,I)$. 
It means that the system is integrable 
and the classical trajectories of the 6-dimensional 
phase-space are restricted to a 2-dimensional torus.$^{7}$  
For a fixed value of $J$ the Hamiltonian (5) 
describes a sort of nonlinear pendulum 
where the extrema of integration are 
\beq 
\theta_1 = arcos{\left[ \sqrt{2H-aJ^2\over 2 (b-a)} \right] }  \; , 
\;\;\;\;\;\;\;\; 
\theta_2 = arcos{\left[ -\sqrt{2H-aJ^2\over 2 (b-a)} \right] }  \; , 
\eeq 
for $0 \le H < (b/2)J^2$ and it corresponds to librational motion, 
while they are 
\beq 
\theta_1 =0 \; , \;\;\;\;\;\;\;\; \theta_2 = \pi \; , 
\eeq 
for $(b/2) J^2 < H \leq (c/2) J^2$ and it corresponds 
to rotational motion. Note that $H=(b/2) J^2$ is the energy of 
the separatrix between librational and rotational motions. 
The semiclassical (torus) quantization$^{11}$ of the energy 
is performed by setting 
\beq 
J = \left (j+{1\over 2} \right ) \hbar  , \;\;\;\; 
I = k \hbar  \; ,  
\eeq 
where $j$ and $k$ are integer quantum numbers 
such that $k=-j,-j+1,...,j-1,j$ and $j\geq 0$. 
Note that only recently the corrections to this torus 
quantization, namely the higher order terms of the WKB 
series of the angular momentum, have been investigated.$^{12}$  
For a fixed value of $j$ there is a set of $2j+1$ energy levels 
but the semiclassical energy spectrum $E_{j,k}^{sc}$, given by the 
implicit equation 
\beq 
k\hbar = {1\over \pi} \left (j+{1\over 2} \right ) \hbar 
\int_{\theta_1}^{\theta_2}  
\sqrt{ {2 \epsilon_{j,k}^{sc}} 
- a -(b-a)\cos^2{\theta} 
\over (c-a) - (b-a) \cos^2{\theta} } \; d\theta \; ,  
\eeq 
where $\epsilon_{j,k}^{sc}:= E_{j,k}^{sc} /[(j+{1/2})^2 \hbar^2]$,
has $j$ couples of degenerate levels 
because $E_{j,k}^{sc}=E_{j,-k}^{sc}$. 
The lowest energy level of the set is 
$E_{j,0}^{sc}=a \hbar^2 (j+1/2)^2/2$ while the higher 
is $E_{j,j}^{sc}=c\hbar^2 (j+1/2)^2/2$. 

\section{Numerical results: spectral fluctuations} 

In Figure 1 we compare the "exact" energy spectrum, obtained 
with the numerical diagonalization$^{6}$ 
of the matrix Hamiltonian given by Eq. (2) and Eq. (3), with 
the semiclassical one, calculated using Eq. (10). 
As expected, the semiclassical quantization cannot take into account 
the broken degeneracy of pairs of ``exact'' energy levels. 
Moreover we find that the semiclassical energy levels are more 
accurate in the higher part of the spectrum (rotational levels). 
\par
Berry and Tabor$^{1,2}$ have shown that for classically integrable 
systems the spectral statistics $P(s)$ and $\Delta_3(L)$ 
are expected to follow Poisson predictions: 
$P(s)=\exp{(-s)}$ and $\Delta_3(L)= {L\over 15}$. 
These predictions are based on the analysis of the semiclassical 
energy spectrum of a generic integrable Hamiltonian with more than 
one action variable. 
Thus, to test the Berry-Tabor theory, is much more appropriate 
to investigate spectral statistics of the semiclassical energy 
levels than spectral statistics of ``exact'' energy levels, 
as done in Ref. 6. 
\par 
In Figure 2 we plot $P(s)$ and $\Delta_3(L)$ obtained from the 
semiclassical energy spectrum. The level spectrum is mapped into 
unfolded levels with quasi-uniform level density by using a standard 
procedure described in Ref. 13. 
$P(s)$ is the distribution of nearest-neighbor spacings 
$s_i=({\tilde E}_{i+1}-{\tilde E}_i)$ 
of the unfolded levels ${\tilde E}_i$. 
It is obtained by accumulating the number of spacings that lie within 
the bin $(s,s+\Delta s)$ and then normalizing $P(s)$ to unit. 
The statistic $\Delta_3(L)$ is defined, for a fixed interval 
$(-L/2,L/2)$, as the least-square deviation of the staircase 
function $N(E)$ from the best straight line fitting it,   
where $N(E)$ is the number of levels between E and zero for positive 
energy, between $-E$ and zero for negative energy. 
Figure 3 shows that the spectral distribution $P(s)$ has a 
peak near s=1 and nothing elsewhere. The spectral rigidity $\Delta_3(L)$ 
follows the Poisson prediction $\Delta_3(L)=L/15$ 
for small $L$ but for larger values of $L$ it gets a constant 
mean value with fluctuations around this mean value. 
\par 
The results of Figure 2 and Figure 3 can be explained 
in the following way. As shown in the previous section, 
fixing the angular momentum $J$ the triaxial rigid 
rotator is one-dimensional: in the case of systems with only 
one action variable the energy spectrum is locally harmonic and, 
after unfolding, the spectral fluctuations have the pathological 
behavior of the one-dimensional harmonic oscillator, thus 
the spacings are all close to one. 
Moreover, it is important to observe 
that both ``exact'' and semiclassical levels have 
a linear grow apart near the separatrix where they cluster. 
This effect is shown in Figure 3 and Figure 4 
where we plot the true nearest neighbor 
spacings (without unfolding) of 
semiclassical and ``exact'' energy levels as a function 
of the energy for a fixed value of the angular momentum. 
The spacings of the semiclassical energy 
levels decrease quasi-linearly with the energy 
up to the separatix and then they increase quasi-linearly. 
The same behavior is found for the spacings of the ``exact'' energy 
levels. In particular, in Figure 4 we plot the spacings 
of the exact energy levels corresponding to the four classes 
of symmetry of the system: the total Hamiltonian matrix 
is decomposed in the direct product of $4$ submatrices 
by considering the parity of the quantum 
number $k$: even (E) or odd (O), and the symmetry of the state: 
symmetric (S) or anti-symmetric (A). 
So the submatrices are labelled as follow: (E,S), (E,A), (O,S), (O,A). 
These are the classes of symmetry of the system.$^{6,9}$ 
Note that the spacings of each class of symmetry have the 
same trend but there are some fluctuations, in particular 
around the extrema of the energy spectrum. 
We have verified that these results do not depend on the choice 
of the inertia parameters of the asymmetric rigid rotator. 

\section{Algebraic Harmonic Approximation close to minima 
and maxima: single levels} 

The investigation of the triaxial-top classical Hamiltonian with 
$a<b<c$ reveals that the minimum-energy configuration 
is doubly realized by $(J_1,J_2,J_3)=(\pm J,0,0)$, where $J$ is the
radius of the sphere $J^2_1 +J^2_2 +J^2_3 = J^2$ and represents a 
constant of motion. 
On the other hand, the configurations $(J_1,J_2,J_3)=(0,0,\pm J)$ 
and $(J_1,J_2,J_3)=(0,\pm J, 0)$ are recognized to be  
maxima and saddle points, respectively. By using the
Casimir operator ${\hat J}^2$, we notice that 
the quantum Hamiltonian $\hat H$ can 
be rewritten in the two forms
\beq
\hat H =
\frac{_1}{^2} \left [ 
a {\hat J}^2 + (b - a ) {\hat J}_2^2 +(c - a ) {\hat J}_3^2 \right ] \; , 
\eeq
\beq
\hat H =  
\frac{_1}{^2} \left [ c {\hat J}^2 - (c -a ) {\hat J}_1^2 -  
(c -b) {\hat J}_2^2 \right ] \; , 
\eeq
close to the minima and the maxima, respectively. 
The observation that, classically, the energy minimum
is reached for $J_2 = J_3 =0$, and $J_1 = \pm J$, leads to 
approximate $J_1$ as $J_1 \simeq 
\pm J[ 1 - (J^2_2 +J^2_3)/2J^2]$. This implies, 
at the quantum level, the inequalities 
$\langle {\hat J}_s \rangle \ll J$ [where the radius $J$ is 
given quantally by $J^2 = j(j+1)\hbar$] concerning
the expectation values of ${\hat J}_s$ ($s= 2,\, 3$) and makes natural to 
adopt the perturbative scheme (see Ref. 14) in which the 
operator ${\hat J}_2$ and ${\hat J}_3$ can be treated as 
canonically conjugate variables. Such a feature follows from 
\beq
[{\hat J}_2, {\hat J}_3] \simeq\, i J \, ,
\label{COM}
\eeq
wheras 
\beq
[{\hat J}_3, {\hat J}_1] = i J ({\hat J}_2/ J) \simeq 0 \; , \;
[{\hat J}_1, {\hat J}_2] = i J ({\hat J}_3/ J) \simeq 0 \, ,
\eeq
are considered vanishing when compared with (\ref{COM}) since
$\langle {\hat J}_s \rangle / J << 1$. Then, after labeling by
the integer $k$ the eigenstates of the harmonic oscillator term
$(c-a ) [{\hat J}_3^2+ W^2 {\hat J}_2^2 ]$ contained in $\hat H$,
where $W^2= (b -a )/(c-a )$, its eigenvalues are found to be
$J W (2k +1)$ so that the spectrum of $\hat H$ reads
\beq 
E_{j,k}^{har} = \frac{_1}{^2} \left [
a j(j+1) \hbar^2 + \sqrt{j(j+1)} \hbar 
\sqrt{(b -a )(c-a )} \, (2k+1) \hbar \right ] \; .
\label{EIG1} 
\eeq 
The approximation just implemented is not consistent 
with the spectrum expected for $c \to a$. In fact, 
in such a case the Hamiltonian becomes 
${\hat H} = [ c J^2 +(b-a ) {\hat J}_1^2 ]/2$ 
thus entailing
however a quadratic spectrum $E = [ c J^2 + (b - a) k^2\hbar^2 ]/2$. 
Similarly, the approximation for the maxima is obtained by
noticing that 
$J_3 \simeq \pm J[ 1 -( {\hat J}^2_1 + {\hat J}_2^2)/2J^2]$, and
\beq
[{\hat J}_1, {\hat J}_2] \simeq\, i J \, ,
\label{TOM}
\eeq
\beq
[{\hat J}_3, {\hat J}_1] = i J ({\hat J}_2/J) \simeq 0 \; , \;
[{\hat J}_2, {\hat J}_3] = i J ({\hat J}_1/J) \simeq 0 \, .
\eeq
In this case, the harmonic oscillator term 
$(c-b) [{\hat J}_2^2 + \Omega^2 {\hat J}_1^2]$, with 
$\Omega^2 = (c-a )/(c-b )$, contained in $\hat H$ leads to the spectrum
\beq
E_{j,k}^{har} = 
\frac{_1}{^2} \left [
c j(j+1) \hbar^2 - \sqrt{j(j+1)} \hbar \sqrt{(c -a )(c -b )} 
\, (2k+1)\hbar \right ] \; ,
\label{EIG2}
\eeq
which differs from Eq. (\ref{EIG1}) owing to the interlevel separation. 
In Figure 5 we compare "exact" energy levels with those obtained 
using the harmonic approximation. 
The results show that indeed the harmonic approximation is quite accurate, 
also better than the semiclassical one, close to the minimum and the 
maximum of the energy. 

\section*{Conclusions} 

In this paper we have investigated the energy spectrum of the 
triaxial rigid rotator, a very important model for both atomic and 
nuclear physics. We have found that its 
semiclassical energy spectrum is spanned by two quantum numbers: 
the quantum number $j$ of the total angular momentum $J$ 
and the quantum number $k$ of the other action variable $I$ 
of the system. For a fixed value of $j$ 
the semiclassical energy spectrum depends on the quantum 
number $k$ and can be divided into librational 
and rotational energy levels. Moreover, the spacings of both 
``exact'' and semiclassical energy spectra show a quasi-linear 
behavior as a function of energy: nearest neighbour 
spacings have a quasi-linear decrease up to the energy of the separatix 
and then a quasi-linear increase. 
We have also shown that the semiclassical energy spectrum 
is in good agreement with the ``exact'' one, but cannot  
take into account the broken degeneracies of pairs of ``exact'' 
energy levels. From our analysis it follows that 
the spectral fluctuations of the asymmetric top 
have the pathological behavior of the one-dimensional 
harmonic oscillator. This effect has been verified by 
calculating the spectral statistics $P(s)$ and $\Delta_3(L)$. 
Finally, we have deduced analytical formulas, 
based on an algebric approximation of the quantum Hamiltonian, 
which describe remarkably well the energy levels of the triaxial 
rotor close to extrema of the energy. 
\par
V.R.M. is greately indebted to M.V. Berry, S. Graffi and A. Turchetti 
for many suggestions at different stages of the paper. 
L.S. thanks T. Prosen for enlightening discussions. 

\section*{References}

\begin{enumerate}

\item{} M.V. Berry and M. Tabor, 
Proc. Roy. Soc. Lond. A {\bf 356}, 375 (1977); 
M.V. Berry, Annals of Phys. {\bf 131}, 163 (1981); 
M.V. Berry, Proc. Roy. Soc. Lond. A {\bf 400} 229 (1985). 

\item{} M. Tabor, 
{\it Chaos and Integrability in Nonlinear Dynamics} 
(Wiley, New York, 1989). 

\item{} V.R. Manfredi and L. Salasnich,    
Int. J. Mod. Phys. B {\bf 13}, 2343 (1999). 

\item{} A. Pandey, O. Bohigas and M.J. Giannoni,  
J. Phys. A: Math. Gen. {\bf 22} 4083 (1989). 

\item{} A. Pandey and R. Ramaswamy, Phys. Rev. A 
{\bf 43}, 4237 (1991). 

\item{} V.R. Manfredi and L. Salasnich, 
Phys. Rev. E {\bf 64}, 066201 (2001). 

\item{} V.I. Arnold, {\it Mathematical Methods in Classical 
Mechanics} (Springer, New York, 1984).  

\item{} A.R.P. Rau, Rev. Mod. Phys. {\bf 64} 623 (1992);  
A. Turchetti, {\it Dinamica Classica dei Sistemi Fisici} 
(Zanichelli, Bologna, 1998) 

\item{} L. Landau and E. Lifshitz, {\it Course in Theoretical Physics},  
vol. 3: Quantum Mechanics (Pergamon, London, 1977).  

\item{} J. Villain, J. de Phys. {\bf 35}, 27 (1974). 

\item{} V.P. Maslov and M.V. Fedoriuk, 
{\it Semiclassical Approximation 
in Quantum Mechanics} (Reidel Publishing Company, New York, 1981)

\item{} M. Robnik and L. Salasnich, J. Phys. A (Math. Gen.) 
{\bf 30}, 1719 (1997); 
L. Salasnich and F. Sattin, J. Phys. A (Math. Gen.) {\bf 30},  
7597 (1997); M. Robnik and L. Salasnich, Nonlinear Phenomena 
in Complex Systems {\bf 3}, 99 (2000). 

\item{} V.R. Manfredi, Lett. Nuovo Cimento {\bf 40}, 135 (1984). 

\item{} R. Franzosi, V. Penna and R. Zecchina, 
Int. J. Mod. Phys. B {\bf 14}, 943 (2000). 

\end{enumerate}

\newpage

\begin{figure}
\centerline{\psfig{file=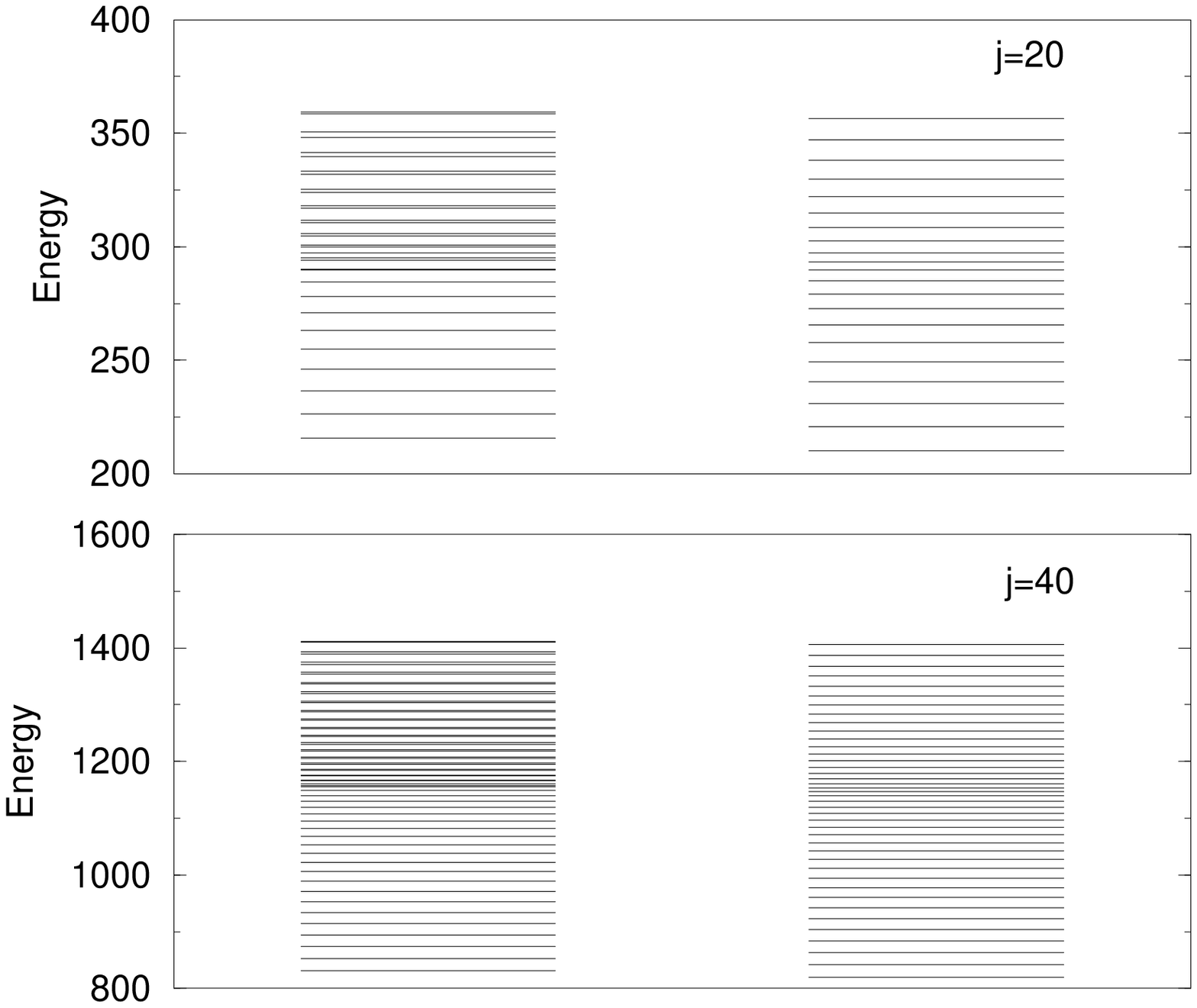,height=5.in}}
\vskip 1.cm
%\caption
{Fig. 1: ``Exact'' (left) vs semiclassical (right) 
energy spectrum of the triaxial rigid rotator with a fixed value 
of the angular momentum $j$. Parameters: $a=1$, $b=\sqrt{2}$, 
$c=\sqrt{3}$ and $\hbar=1$. Energy in arbitrary units.} 
\end{figure}

\newpage

\begin{figure}
\centerline{\psfig{file=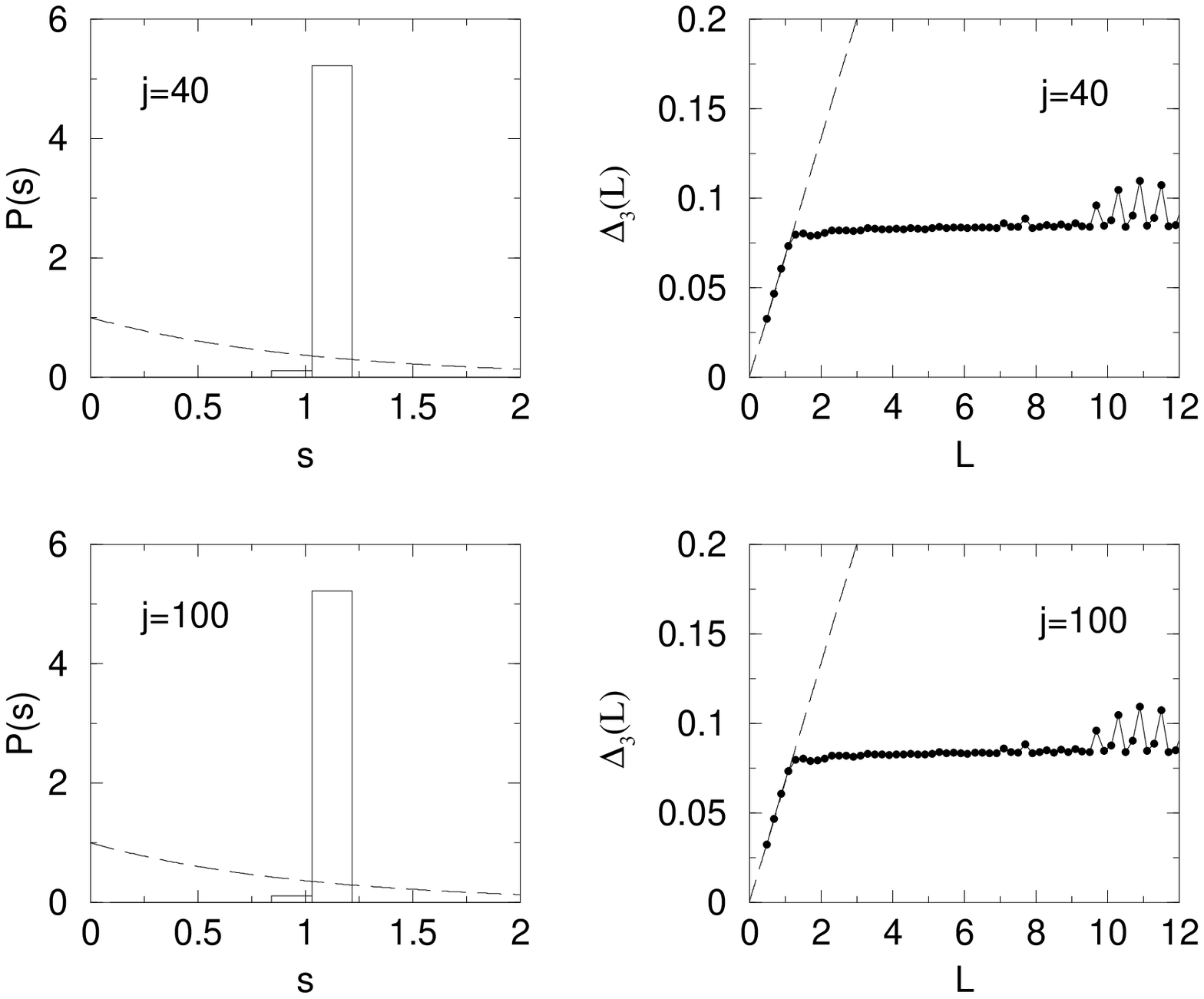,height=5.in}}
\vskip 1.cm 
%\caption
{Fig. 2: Nearest neighbor spacing distribution 
$P(s)$ and spectral rigidity $\Delta_3(L)$ of the semiclassical 
energy spectrum of the triaxial rigid rotator with a fixed 
value of the angular momentum $j$. 
The dashed lines are the Poisson predictions: 
$P(s)=\exp{(-s)}$ and $\Delta_3(L)=L/15$. 
Parameters as in Figure 1} 
\end{figure}

\newpage

\begin{figure}
\centerline{\psfig{file=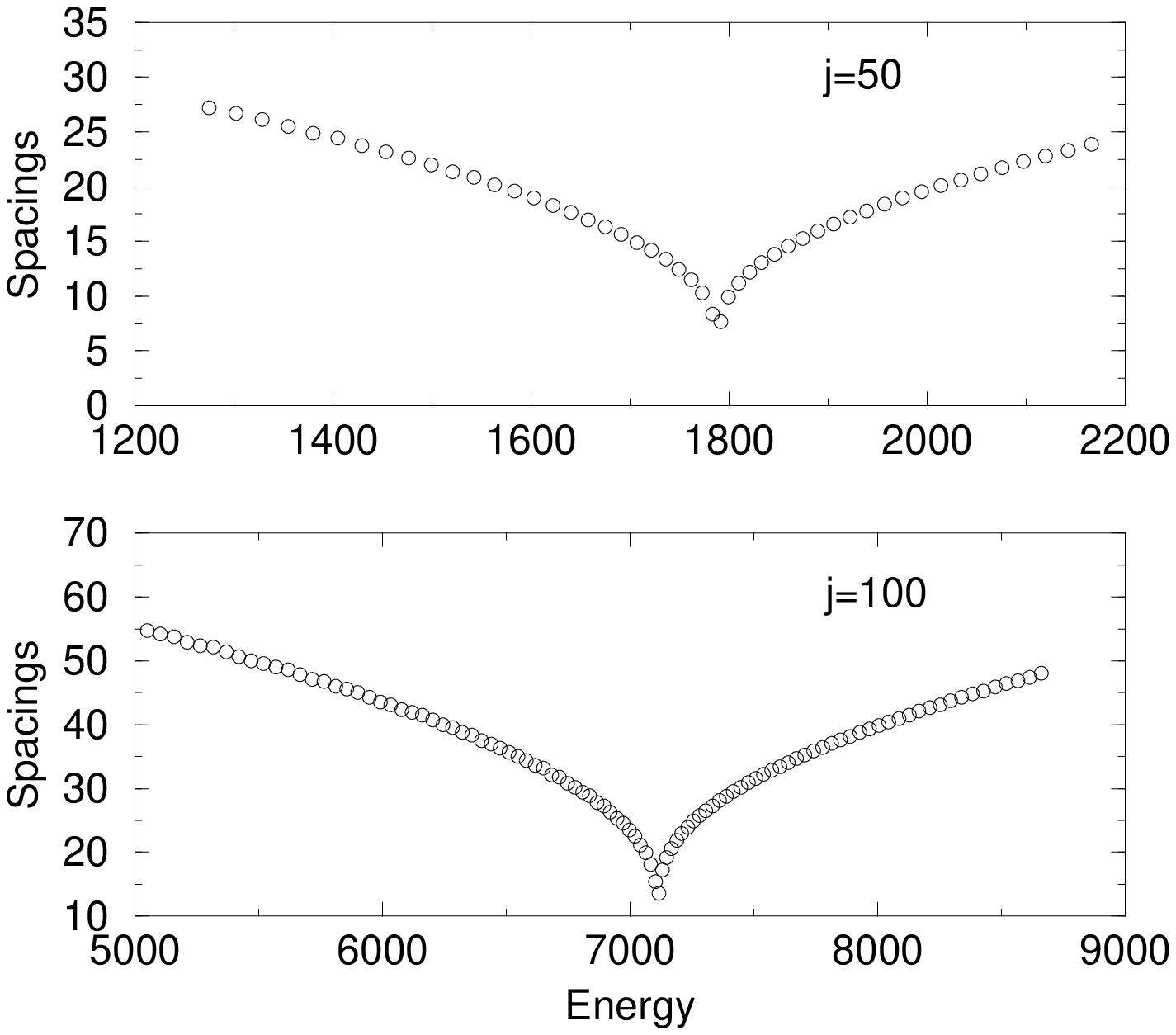,height=5.in}}
\vskip 1.cm
%\caption
{Fig. 3: True spacings (without unfolding) as a function of energy. 
Semiclassical energy levels of the triaxial rigid rotator 
for two values of the angular momentum $j$. 
Parameters and units as in Figure 1.} 
\end{figure}

\newpage

\begin{figure}
\centerline{\psfig{file=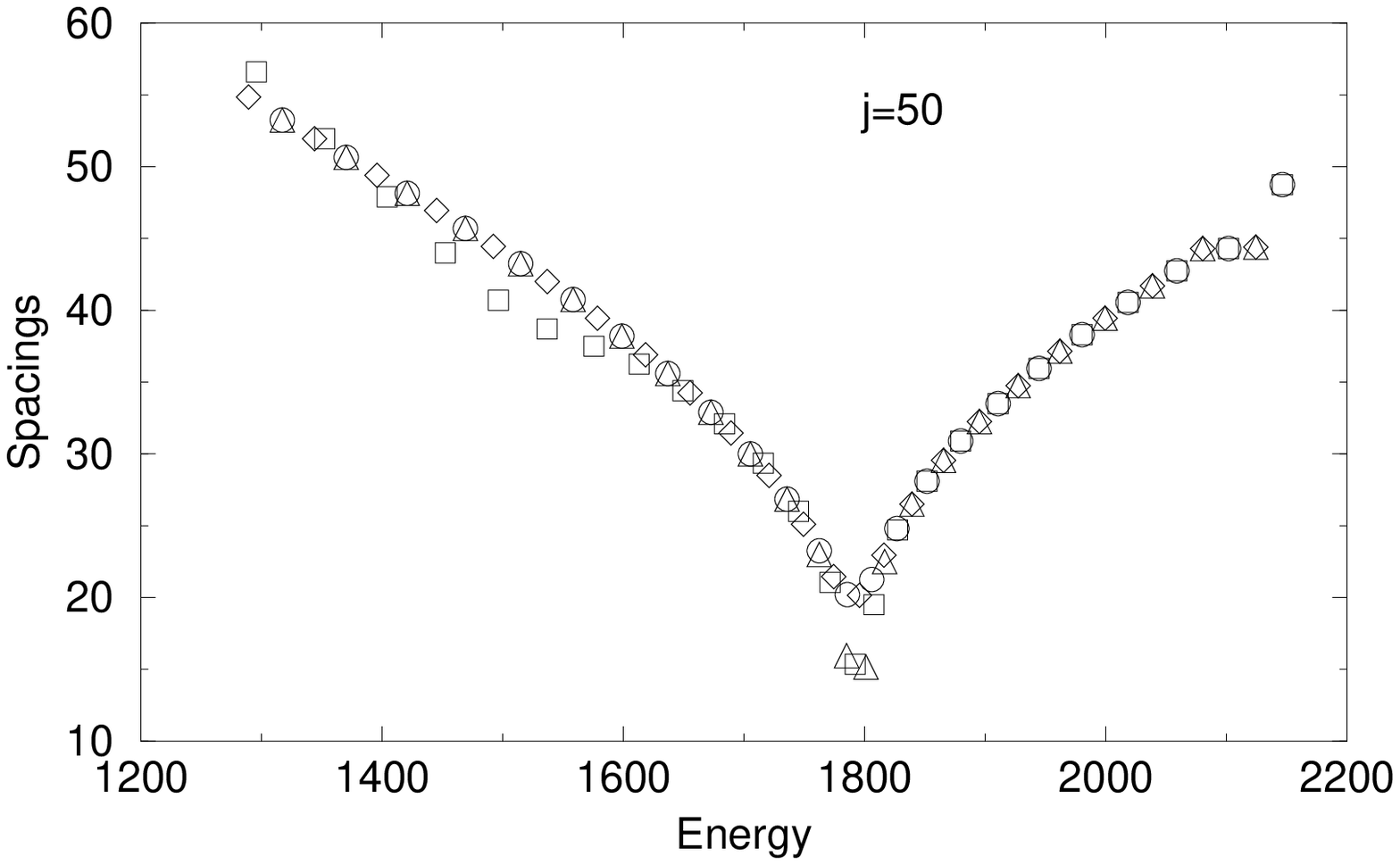,height=3.5in}}
\vskip 1.cm
%\caption
{Fig. 4: True spacings (without unfolding) as a function of energy. 
``Exact'' energy levels of the triaxial rigid rotator 
for a fixed value of the angular momentum $j$. 
Four classes of symmetry: square for class (E,S), circle for 
class (E,A), diamond for class (O,S), triangle for class (O,S). 
Parameters and units as in Figure 1.} 
\end{figure} 

\newpage

\begin{figure}
\centerline{\psfig{file=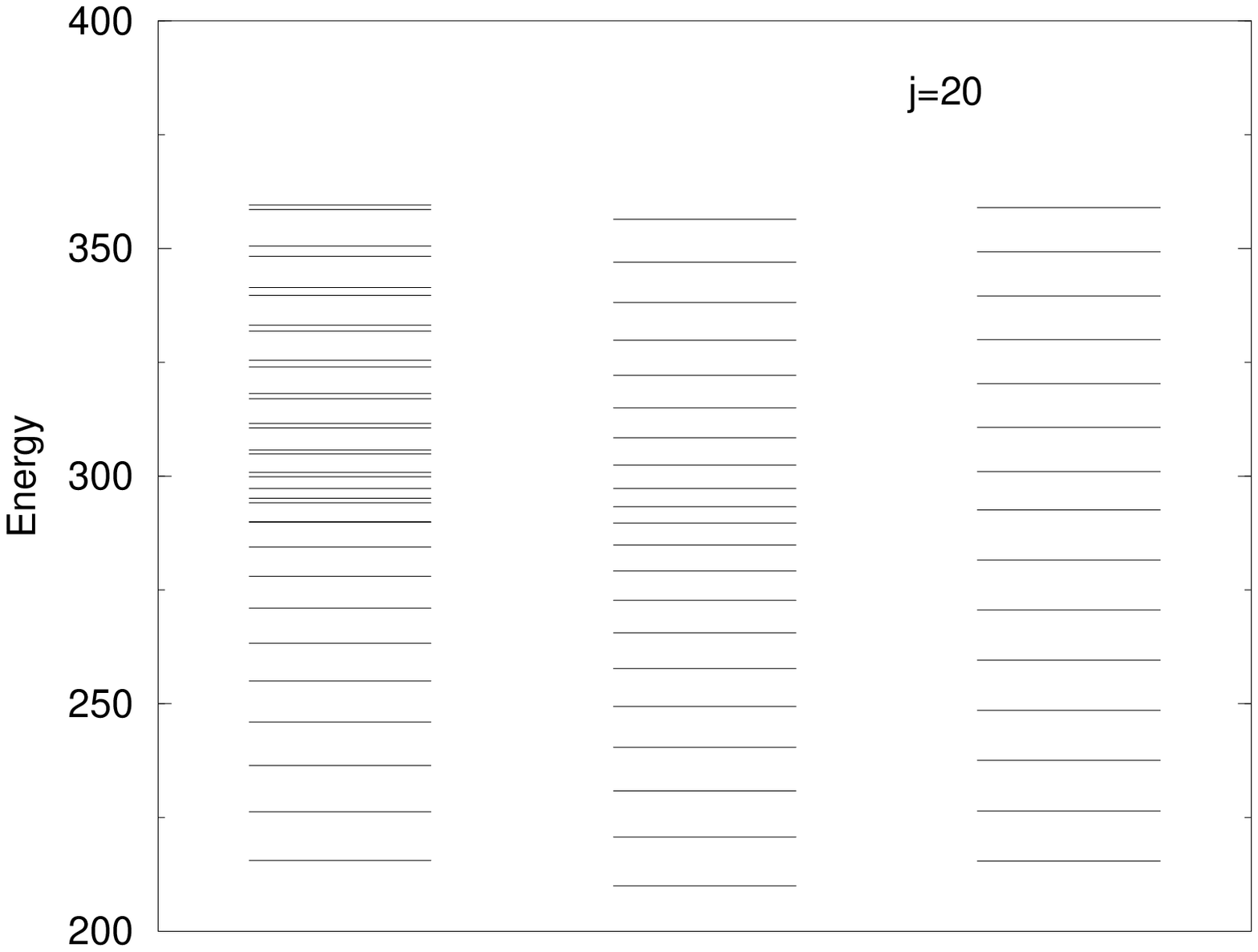,height=4.in}}
\vskip 1.cm 
%\caption
{Fig. 5: ``Exact'' (left), semiclassical (middle) and 
harmonic-approximation (right) energy spectrum of the triaxial 
rigid rotator with a fixed value of the angular 
momentum $j$. Parameters and units as 
in Figure 1.} 
\end{figure} 

\end{document}